\documentclass[a4paper, 8pt]{article}
\usepackage{amsmath}
\usepackage{amsthm}
\usepackage{amsfonts}
\usepackage{amssymb}
\usepackage{mathtools}
\usepackage{bbm}
\usepackage{graphicx}
\usepackage{lscape}
\usepackage{booktabs}
\usepackage{dsfont}
\usepackage{hyperref}
\usepackage{bm}
\usepackage{fullpage}
\usepackage{xcolor}
\usepackage[english]{babel}
\usepackage[applemac]{inputenc}
\usepackage{soul}
\usepackage{cancel}
\usepackage[titletoc,toc,title]{appendix}
\usepackage{authblk}

\title{Rainbow metric from quantum gravity: anisotropic cosmology}
\author[1]{Mehdi Assanioussi}
\author[2]{Andrea Dapor}
\date{}
\affil[1]{Faculty of Physics, University of Warsaw\\

Poland\\

\url{mehdi.assanioussi@fuw.edu.pl}}

\affil[2]{Institute for Quantum Gravity, Friedrich-Alexander University Erlangen-N¸rnberg\\

Germany\\

\url{andrea.dapor@gravity.fau.de}}

\begin{document}

\maketitle
\begin{abstract}
In this paper we present a construction of effective cosmological models which describe the propagation of a massive quantum scalar field on a quantum anisotropic cosmological spacetime. Each obtained effective model is represented by a rainbow metric in which particles of distinct momenta propagate on different classical geometries. Our analysis shows that upon certain assumptions and conditions on the parameters determining such anisotropic models, we surprisingly obtain a unique deformation parameter $\beta$ in the modified dispersion relation of the modes. Hence inducing an isotropic deformation despite the general starting considerations. We then ensure the recovery of the dispersion relation realized in the isotropic case, studied in \cite{rainbow1}, when some proper symmetry constraints are imposed, and we estimate the value of the deformation parameter for this case in loop quantum cosmology context.
\end{abstract}
\vspace{0,5cm}
How to recover classical spacetime from a fundamentally quantum description of geometry is a long-standing question in quantum gravity. A promising idea is that classical gravity could be a collective phenomenon emerging from quantum degrees of freedom \cite{emerg1,emerg2}, not unlike fluid dynamics emerges from microscopic molecular interactions. Taking a pragmatic point of view, we note that what an observer really measures is matter, not geometry: through matter propagation, she \emph{infers} the geometry. Thus, given a certain dynamics for the matter content, every geometry which is consistent with such dynamics is equally good. In light of this fact, in \cite{QT-FRW} the authors derived the dynamics of a quantum scalar field (the matter) propagating on a quantum cosmological spacetime (the geometry), and looked for classical spacetimes which would produce the same dynamics for such a scalar field. It turns out that a possible \emph{effective spacetime} exists, whose metric is given by certain expectation values of geometric operators on the quantum state of geometry (for this reason, it was called ``dressed metric''). This fact -- i.e., the possibility of giving an equivalent description of QFT on quantum spacetime in terms of QFT on a classical spacetime -- simplifed the treatment of quantum spacetime in several scenarios, such as pre-inflationary cosmological perturbations \cite{QT-pert}, particle creation in primordial cosmology \cite{QT-particles} and Hawking radiation from quantum spherical black holes \cite{QT-BH}.

From a purely theoretical perspective, however, we must point out that the effective spacetime proposed in \cite{QT-FRW} is not unique, unless the scalar field is free and massless. In particular, in \cite{rainbow1} we focused on the massive free scalar field case, finding an alternative dressed metric for the same underlying quantum system. The peculiarity of this result lies in the fact that such metric depends on the energy of the field quanta under consideration (despite having explicitly made use of the test-field approximation \cite{alex}, in which one disregards the backreaction of matter on geometry), which in turn leads to an apparent Lorentz-symmetry violation. This was confirmed in \cite{liberati}, where the authors studied the dispersion relation for the scalar field on such a quantum spacetime.

The conclusion is that particles of different energy probe ``different classical aspects'' of the same quantum state of geometry. Although the fundamental quantum system may be Lorentz-invariant, an observer measuring the propagation of particles could equally well give a description in terms of a QFT on an energy-dependent metric (known in the literature as ``rainbow metric" \cite{RAIN-intro1,RAIN-intro2}). But this is by no means surprising, in light of the emergent spacetime concept. Indeed, Lorentz-violating effects are common in condensed matter physics, e.g., in the propagation of light through crystals.

While very compelling, this result remains nevertheless limited to the case in which the quantum geometry is homogeneous and isotropic. As such, it is perhaps not surprising to find out that the modification to the dispersion relation is minimal, amounting to a redefinition of the speed of light by a parameter of quantum gravity origin which can only depend on time. It has in fact been argued that, to see more interesting (and potentially measurable) deviations, one needs to take into account more degrees of freedom for the quantum geometry. This is the purpose of a long-term project: we will take a bottom-up approach in generalizing the concept of energy-dependent dressed metric, starting in the current work by lifting the isotropy requirement. Specifically, in this paper we apply the same construction of \cite{rainbow1} to the Bianchi I case. As we will see, the deformation of the dispersion relation is again described by a single parameter (i.e, it is an isotropic deformation like in the FLRW case), but such parameter depends non-trivally on the quantum anisotropies. This is nevertheless consistent with the isotropic case studied in \cite{rainbow1} when the symmetry reduction is realized.

The structure of the paper is the following. In Section 1, we review the Hamiltonian formulation of the classical theory, emphasizing the canonical coordinates to be quantized and the choice of time (which is necessary in order to obtain a true physical dynamics). In Section 2, the quantization of the theory is performed (note that, for all intents and purposes, this step is completely general: in no way we are limiting ourselves to a specific theory of quantum gravity), and a system of equations is found, whose solutions correspond to compatible dressed metrics. Section 3 is dedicated to the analysis of dispersion relations: following \cite{liberati}, we observe that it is not necessary to know the complete dressed metric solution in order to extract the \emph{exact} dispersion relation for the scalar field (in fact, only the low-energy limit is required). This intuition allows us to find the explicit form of the Lorentz-deformed dispersion relation in the Bianchi case which -- as already said -- turns out to be controlled by a unique parameter. We also notice that this parameter is compatible with the FLRW case. In Section 4, we estimate the deformation parameter for a semiclassical state of geometry in isotropic loop quantum cosmology, finding that it is proportional to the square of the spread, and hence extremely small for peaked states which puts it well within the experimental bounds discussed in \cite{liberati}. Finally, in Section 5 we summarize the results and comment on future developments.
		\section{Classical construction}
The system we want to study is a test massive scalar field (denoted by $\phi$) propagating on a class of anisotropic spacetimes of the Bianchi I type. We also include a homogeneous massless scalar field $\Phi$ as the source, which will play the role of relational time. The action of the model is therefore $S = S_G[g] + S_\Phi[\Phi, g] + S_M[\phi, g]$, where $S_G$ is Einstein-Hilbert action and
\begin{align}
S_\Phi[\Phi, g] &= \int d^4x \ \mathcal{L}_\Phi = -\dfrac{1}{2} \int d^4x \sqrt{-g} g^{\mu\nu} \partial_\mu \Phi \partial_\nu \Phi
\\
S_M[\phi, g] &= \int d^4x \ \mathcal{L}_M = -\dfrac{1}{2} \int d^4x \sqrt{-g} \left(g^{\mu\nu} \partial_\mu \phi \partial_\nu \phi + m^2 \phi^2\right)
\end{align}
The metric for Bianchi I class of spacetimes is
\begin{align}
ds^2 = g_{\mu\nu} dx^\mu dx^\nu = -N^2 dT^2 + \sum_{i = 1}^3 a_i^2(t) (dx^i)^2
\end{align}
where $N$ is the lapse function corresponding to the time coordinate $T$ and $a_i$ are the scale factors along the three spatial directions. Plugging this metric in $S_\Phi$ and $S_M$, and using homogeneity of $\Phi$, one finds the simplified versions
\begin{align}
S_\Phi = \dfrac{1}{2} \int dT \dfrac{a_1a_2a_3}{N} \dot{\Phi}^2, \qquad S_M = -\dfrac{1}{2} \int dT d^3x \ N a_1a_2a_3 \left(-\dfrac{\dot{\phi}^2}{N^2} + \sum_i \dfrac{(\partial_i \phi)^2}{a_i^2} + m^2 \phi^2\right)
\end{align}
where the dot represents derivative with respect to $T$. In order to obtain the Hamiltonian, we perform a Legendre transform. First, identify the momentum $\pi$ conjugated to $\phi$:
\begin{align}
\pi := \dfrac{\delta S_M}{\delta \dot \phi} = \dfrac{a_1a_2a_3}{N} \dot{\phi}
\end{align}
The Hamiltonian for $\phi$ is then given in terms of the Lagrangian $\mathcal{L}_M$ by
\begin{align}
\nonumber H_M & = \int d^3x \left(\pi \dot{\phi} - \mathcal{L}_M\right) = \dfrac{1}{2} \dfrac{N}{a_1a_2a_3} \int d^3x \left(\pi^2 + (a_1a_2a_3)^2 \left(\sum_i \dfrac{(\partial_i \phi)^2}{a_i^2} + m^2 \phi^2\right)\right)\\
& = \dfrac{1}{2} \dfrac{N}{\sqrt{p_1p_2p_3}} \int d^3x \left(\pi^2 + \sum_i (p_i \partial_i \phi)^2 + p_1p_2p_3 \ m^2 \phi^2\right)
\end{align}
where in the last step we defined $p_1 := a_2 a_3$ and $p_2$ and $p_3$ cyclically. By performing a Fourier expansion of $\phi$ and $\pi$ we can finally obtain an expression in terms of wave vectors $\vec{k}$ (taking values in a 3-dimensional lattice if we assume the space topology to be that of a torus):
\begin{align} \label{matter-H-class}
H_M = \sum_{\vec{k}} H_{\vec{k}} = \dfrac{1}{2} \dfrac{N}{\sqrt{p_1p_2p_3}} \sum_{\vec{k}} \left[\pi_{\vec{k}}^2 + \left(\sum_i (p_i k_i)^2 + p_1p_2p_3 \ m^2\right) \phi_{\vec{k}}^2\right]
\end{align}
This expression is valid for every choice of time coordinate $T$. A particularly interesting choice is given in terms of the homogeneous field $\Phi$, because in this case the time coordinate acquires physical meaning. Since for such a scalar field the momentum is
\begin{align}
\Pi := \dfrac{\delta S_\Phi}{\delta \dot \Phi} = \dfrac{\sqrt{p_1p_2p_3}}{N} \dot{\Phi}
\end{align}
the Hamiltonian for $\Phi$ is $H_\Phi = \int dT \ N \Pi^2/\sqrt{p_1p_2p_3}$, contributing to the scalar constraint:\footnote
{
While $\phi$ is a test field, $\Phi$ is the source of the geometry, and hence plays a role in defining the evolution of $p_i$. 
}
\begin{align} \label{ham-con}
C[N] = \frac{1}{2} \int d^3x \ \dfrac{N}{\sqrt{p_1p_2p_3}}\ \Pi^2 + C_G[N]
\end{align}
where $C_G[N] = \int d^3x \ NC_G$ is the gravitational part of the scalar constraint. This constraint is linear in $\Pi$ if we choose $N = \sqrt{p_1p_2p_3}/\Pi$, and equation $C[N] = 0$ can then be solved for $\Pi$:
\begin{align}
\Pi = \sqrt{-2\sqrt{p_1p_2p_3} C_G}
\end{align}
This means that the specific choice of foliation identified by the lapse function
\begin{align} \label{physical-lapse}
N = \dfrac{\sqrt{p_1p_2p_3}}{\Pi}
\end{align}
corresponds to the choice of $\Phi$ as time coordinate. The $\Phi$-evolution of every phase space function $F$ (that is, function of the gravitational degrees of freedom, $p_i$, and their momenta) is then given by
\begin{align}
\dfrac{dF}{d\Phi} = \{F, \Pi\} = \{F, \sqrt{-2\sqrt{p_1p_2p_3} C_G}\}
\end{align}
so $\Pi$ has the meaning of physical Hamiltonian for the background spacetime. We denote it by $H_o$ to distinguish it from the matter Hamiltonian $H_M$, which for the choice (\ref{physical-lapse}) becomes
\begin{align}
H_M = \dfrac{1}{2\Pi} \sum_{\vec{k}} \left[\pi_{\vec{k}}^2 + \left(\sum_i (p_i k_i)^2 + p_1p_2p_3 \ m^2\right) \phi_{\vec{k}}^2\right]
\end{align}
and defines the $\Phi$-evolution of the test field $\phi$.
		\section{Quantization and effective models}
				\subsection{Dressed metric}
The quantization of the system ``gravity plus matter'' is easily performed. Let $\mathcal{H}_G$ and $\mathcal{H}_M$ be the Hilbert spaces of gravitational and matter degrees of freedom respectively.\footnote
{
From now on we will call ``matter'' the test field $\phi$ only, since the field $\Phi$ is not a dynamical variable (though it features as time variable), having dropped from the system when we solved the scalar constraint at the classical level.
}
The full Hamiltonian of the system is implemented on $\mathcal{H}_G \otimes \mathcal{H}_M$ as an operator formally given by
\begin{align}
\hat{H} = \hat{H}_o + \hat{H}_M = \hat{H}_o + \dfrac{1}{2} \sum_{\vec{k}} \left[\widehat{H_o^{-1}} \otimes \hat{\pi}_{\vec{k}}^2 + \left(\sum_i (\widehat{H_o^{-1} p_i^2} \ k_i^2 + \widehat{H_o^{-1} p_1p_2p_3} \ m^2\right) \otimes \hat{\phi}_{\vec{k}}^2\right]
\end{align}
where $\hat{H}_o$ does not depend on matter operators $\hat{\phi}_{\vec{k}}, \hat{\pi}_{\vec{k}}$, and is considered the ``unperturbed Hamiltonian'' of the system. $\hat{H}$ defines the dynamics of any state $\Psi \in \mathcal{H}_G \otimes \mathcal{H}_M$ via the Schroedinger equation
\begin{align} \label{full-ham}
i \dfrac{d}{d\Phi} \Psi(\Phi) = \hat{H} \Psi(\Phi)
\end{align}
In the spirit of test-field approximation, at zeroth order the state has the form of a simple tensor product:
\begin{align}
\Psi(\Phi) = \Psi_o(\Phi) \otimes \varphi(\Phi)
\end{align}
where $\Psi_o \in \mathcal{H}_G$ and $\varphi \in \mathcal{H}_M$. Moreover, since the gravitational part is the background, its evolution is determined by $H_o$ alone, which at the quantum level means that
\begin{align}
i \dfrac{d}{d\Phi} \Psi_o(\Phi) = \hat{H}_o \Psi_o(\Phi)
\end{align}
In light of these observations, we can trace over the gravitational degrees of freedom in (\ref{full-ham}), and are left with the following Schroedinger equation for matter only:
\begin{align} \label{qft-on-qs}
i \dfrac{d}{d\Phi} \varphi(\Phi) = \dfrac{1}{2} \sum_{\vec{k}} \left[\langle \widehat{H_o^{-1}} \rangle \hat{\pi}_{\vec{k}}^2 + \left(\sum_i \langle \widehat{H_o^{-1} p_i^2} \rangle k_i^2 + \langle \widehat{H_o^{-1} p_1p_2p_3} \rangle m^2\right) \hat{\phi}_{\vec{k}}^2\right] \varphi(\Phi)
\end{align}
where expectation values of gravitational operators are taken on state $\Psi_o(\Phi) \in \mathcal{H}_G$.

The important observation is that a similar equation describes the dynamics of a scalar field $\phi$ on a curved (but classical) spacetime of the Bianchi I type. Indeed, looking back at the Hamiltonian in equation (\ref{matter-H-class}), we see that a quantization of the matter degrees of freedom (keeping the geometry classical) leads to the following Schroedinger equation:
\begin{align} \label{qft-on-cs}
i \dfrac{d}{dT} \varphi(T) = \dfrac{1}{2} \sum_{\vec{k}} \dfrac{N}{\sqrt{p_1p_2p_3}} \left[\hat{\pi}_{\vec{k}}^2 + \left(\sum_i (p_i k_i)^2 + p_1p_2p_3 \ m^2\right) \hat{\phi}_{\vec{k}}^2\right] \varphi(T)
\end{align}
By identifying $T$ with $\Phi$, and hence the rhs's of (\ref{qft-on-qs}) and (\ref{qft-on-cs}), we find the following two algebraic equations
\begin{align} \label{easy}
\dfrac{N}{\sqrt{p_1p_2p_3}} = \langle \widehat{H_o^{-1}} \rangle
\end{align}
and
\begin{align}
\dfrac{N}{\sqrt{p_1p_2p_3}} \left(\sum_i (p_i k_i)^2 + p_1p_2p_3 \ m^2\right) = \sum_i \langle \widehat{H_o^{-1} p_i^2} \rangle k_i^2 + \langle \widehat{H_o^{-1} p_1p_2p_3} \rangle m^2
\end{align}
The first one may be immediately solved for $N$, which can be replaced in the second one, leading to a unique equation\footnote{
Notice that there is the possibility of considering, instead of equation (\ref{god}), a system of three equations:
\begin{align}\label{fakegod}
\forall i=1,2,3:\ \qquad p_i^2 k_i^2 + \frac{p_1p_2p_3}{3} \ m^2 - \gamma_i(k_i) = 0
\end{align}
where
\begin{align}
\gamma_i(k_i) := \dfrac{\langle \widehat{H_o^{-1} p_i^2} \rangle}{\langle \widehat{H_o^{-1}} \rangle} k_i^2 + \dfrac{\langle \widehat{H_o^{-1} p_1p_2p_3} \rangle}{3\langle \widehat{H_o^{-1}} \rangle} m^2
\end{align}
It turns out that the solutions to such system are not analytic in $k_i$'s at the $0$ point (See Appendix \ref{proof1} for proof). However, for technical reasons we prefer dealing with effective scale factors that are analytic in the $k_i$'s at low energies, therefore this option is discarded.
}
for the three unknown $p_i$:
\begin{align} \label{god}
\sum_i p_i^2 k_i^2 + p_1p_2p_3 \ m^2 - \gamma(\vec{k}) = 0
\end{align}
where
\begin{align}\label{god2}
\gamma(\vec{k}) := \sum_i \dfrac{\langle \widehat{H_o^{-1} p_i^2} \rangle}{\langle \widehat{H_o^{-1}} \rangle} k_i^2 + \dfrac{\langle \widehat{H_o^{-1} p_1p_2p_3} \rangle}{\langle \widehat{H_o^{-1}} \rangle} m^2
\end{align}
Any solution $\{p_i(\vec{k})\}$ to this equation defines a Bianchi I metric which we call ``effective'' or ``dressed'' metric, because it has the interpretation of the metric seen by a quantum mode $\vec{k}$: the dynamics of this mode on such a classical spacetime is equivalent to that of the same mode on the \emph{quantum} spacetime described by $\Psi_o$.
				\subsection{Alternative method: dressing the mass}
It should be mentioned that solving equation (\ref{god}) is not the only way to find a dressed metric compatible with the underlying quantum system. Indeed, there exists an alternative model based on \cite{QT-pert}, in which one introduces a \emph{renormalized} mass for the effective scalar field. Consider the quantum dynamics of a scalar field $\phi$ of mass $M \neq m$ on a classical Bianchi I spacetime:
\begin{align}
i \dfrac{d}{dT} \varphi(T) = \dfrac{1}{2} \sum_{\vec{k}} \dfrac{N}{\sqrt{p_1p_2p_3}} \left[\hat{\pi}_{\vec{k}}^2 + \left(\sum_i (p_i k_i)^2 + p_1p_2p_3 \ M^2\right) \hat{\phi}_{\vec{k}}^2\right] \varphi(T)
\end{align}
Comparison with (\ref{qft-on-qs}) under the identification of $T$ with $\Phi$ leads to five equations
\begin{align}
\dfrac{N}{\sqrt{p_1p_2p_3}} = \langle \widehat{H_o^{-1}} \rangle, \ \ \ \ \ \dfrac{N}{\sqrt{p_1p_2p_3}} p_i^2 = \langle \widehat{H_o^{-1} p_i^2} \rangle, \ \ \ \ \ N \sqrt{p_1p_2p_3} M^2 = \langle \widehat{H_o^{-1} p_1 p_2 p_3} \rangle m^2
\end{align}
whose unique solution is
\begin{align}
\begin{array}{c}
N = \sqrt[4]{\langle \widehat{H_o^{-1}} \rangle \langle \widehat{H_o^{-1} p_1^2} \rangle \langle \widehat{H_o^{-1} p_2^2} \rangle \langle \widehat{H_o^{-1} p_3^2} \rangle}, \ \ \ \ \ p_i = \sqrt{\dfrac{\langle \widehat{H_o^{-1} p_i^2} \rangle}{\langle \widehat{H_o^{-1}} \rangle}}
\\
M^2 = m^2 \sqrt{\dfrac{\langle \widehat{H_o^{-1}} \rangle \langle \widehat{H_o^{-1} p_1 p_2 p_3} \rangle^2}{\langle \widehat{H_o^{-1} p_1^2} \rangle \langle \widehat{H_o^{-1} p_2^2} \rangle \langle \widehat{H_o^{-1} p_3^2} \rangle}}
\end{array}
\end{align}
In this case the effective metric is therefore the same for all modes, but the effective mass $M$ becomes state-dependent, and hence $\Phi$-dependent. It is interesting to note that the dressing proportionality factor that relates $M$ to $m$ is given by $(1 + \beta)^{-3/4}$ in terms of the parameter $\beta$ (which we introduce below).
		\section{Dispersion relations}\label{section4}
Let us focus on the $\vec{k}$-dependent case, in which the dressed metric is given as a solution of (\ref{god}). Given such a metric $g_{\mu\nu}(\vec{k})$, we can study the (local) dispersion relation of particles propagating on it, and we look in particular for apparent (as opposed to fundamental) violations of local Lorentz symmetry. To do this, we first need to choose a classical observer with respect to which local quantities (i.e., the energy $E$ and the momentum $\vec{P}$ of the particle) are to be computed. If an observer is described herself by a particle in such a spacetime, it is reasonable to consider that it will satisfy the low-energy condition $k_i/m \ll 1$ for all $i$. In this case, we can simplify equation (\ref{god}) to
\begin{align} \label{class-obs}
p^o_1p^o_2p^o_3 = \dfrac{\langle \widehat{H_o^{-1} p_1p_2p_3} \rangle}{\langle \widehat{H_o^{-1}} \rangle}
\end{align}
We have used the superscript ``$o$'' because this is a low-energy expansion of the real solution(s) to (\ref{god}). In other words, we have in general
\begin{align}
p_i(\vec{k}) = p^o_i + O(\vec{k}/m)
\end{align}

Now, suppose a particle of wavevector $k_\mu$ crosses the laboratory of the classical observer. The observer is characterized by its 4-velocity $u^\mu$ and its local spatial frame $e^\mu_i$ ($i = 1, 2, 3$ labels the three spatial vectors used as reference system). If, for simplicity, we consider an orthonormal frame $g^o_{\mu \nu} e^\mu_i e^\nu_j = \delta_{ij}$, then from $g^o_{\mu \nu} u^\mu u^\nu = -1$, $g^o_{\mu \nu} u^\mu e^\nu_i = 0$ we find the coordinate expressions
\begin{align}
u^\mu = (1/N^o, \vec{0}), \ \ \ \ \ e^\mu_1 = (0, 1/a^o_1, 0, 0), \ \ \ \ \ e^\mu_2 = (0, 0, 1/a^o_2, 0), \ \ \ \ \ e^\mu_3 = (0, 0, 0, 1/a^o_3)
\end{align}
Hence, the energy and momentum of the $k$-particle as measured by the observer are
\begin{align}
E := u^\mu k_\mu = \dfrac{k_0}{N^o}, \ \ \ \ \ P_i := e^\mu_i k_\mu = \dfrac{k_i}{a^o_i}
\end{align}
Our purpose is now to write the dispersion relation for the particle of wavevector $k_\mu$ in terms of the physical quantities $E$ and $P_i$. Since it satisfies the mass-shell condition wrt the metric $g_{\mu \nu}(\vec{k})$, we have
\begin{align}
-m^2 & = g^{\mu\nu}(\vec{k}) k_\mu k_\nu = -\dfrac{k_0^2}{N^2} + \sum_i \dfrac{k_i^2}{a_i^2} = -f^2 \left(\dfrac{k_0}{N^o}\right)^2 + g^2 \sum_i \left(\dfrac{k_i}{a^o_i}\right)^2 = \notag
\\
& = -f^2 E^2 + g^2 P^2
\end{align}
where $P^2 = \delta^{ij} P_i P_j$ and we defined
\begin{align}
f^2 := \left(\dfrac{N^o}{N}\right)^2, \ \ \ \ \ \ \ \ \ \ g^2 := \dfrac{\sum_i \left(\dfrac{k_i}{a_i}\right)^2}{\sum_j \left(\dfrac{k_j}{a^o_j}\right)^2}
\end{align}
Note that $g$ can be rewritten using the relation (\ref{easy}), which is valid also for $N^o$ in terms of $p^o_i$:
\begin{align}
g^2 = \left(\dfrac{N^o}{N}\right)^2 \left(\dfrac{N}{N^o}\right)^2 \dfrac{\sum_i \left(\dfrac{k_i}{a_i}\right)^2}{\sum_j \left(\dfrac{k_j}{a^o_j}\right)^2} = f^2 \dfrac{p_1p_2p_3}{p^o_1p^o_2p^o_3} \dfrac{\sum_i \left(\dfrac{k_i}{a_i}\right)^2}{\sum_j \left(\dfrac{k_j}{a^o_j}\right)^2} = f^2 \dfrac{\sum_i (p_i k_i)^2}{\sum_j (p^o_j k_j)^2}
\end{align}
Inverting the mass-shell for $E^2$ we then obtain
\begin{align}
E^2 = \dfrac{1}{f^2} \left(m^2 + g^2 P^2\right) = \dfrac{p_1p_2p_3}{p^o_1p^o_2p^o_3} m^2 + P^2 \dfrac{\sum_i (p_i k_i)^2}{\sum_j (p^o_j k_j)^2}
\end{align}
Now, using equation (\ref{god}) we can replace the term proportional to $m^2$:
\begin{align}
E^2 & = \dfrac{1}{p^o_1p^o_2p^o_3} \left(\gamma(\vec{k}) - \sum_i p_i^2 k_i^2\right) + P^2 \dfrac{\sum_i (p_i k_i)^2}{\sum_j (p^o_j k_j)^2} = \notag
\\
& = \dfrac{1}{p^o_1p^o_2p^o_3} \left(\gamma(\vec{k}) - \sum_i p_i^2 k_i^2\right) + \dfrac{1}{p^o_1p^o_2p^o_3} \sum_i (p_i k_i)^2 = \notag
\\
& = \dfrac{\langle \widehat{H_o^{-1}} \rangle}{\langle \widehat{H_o^{-1} p_1p_2p_3} \rangle} \gamma(\vec{k})
\end{align}
where in the second step we used the expression $P^2 = \sum_i (k_i/a^o_i)^2 = \sum_i (p^o_i k_i)^2/p^o_1p^o_2p^o_3$, and in the last step we replaced $p^o_1p^o_2p^o_3$ with the expectation values according to (\ref{class-obs}). Writing out explicitly $\gamma(\vec{k})$ we finally find the surprisingly simple dispersion relation
\begin{align}
E^2 = m^2 + \sum_i \dfrac{\langle \widehat{H_o^{-1} p_i^2} \rangle}{\langle \widehat{H_o^{-1} p_1p_2p_3} \rangle} k_i^2
\end{align}
To be rigorous, one should write it in terms of physical momentum $P_i$ rather than the wavevector $k_i$. Recalling that $P_i = k_i/a^o_i = k_i p^o_i/\sqrt{p^o_1p^o_2p^o_3}$, we thus get
\begin{align} \label{almost-done}
E^2 & = m^2 + \dfrac{1}{\langle \widehat{H_o^{-1}} \rangle} \sum_i \dfrac{\langle \widehat{H_o^{-1} p_i^2} \rangle}{(p^o_i)^2} P_i^2 = m^2 + P^2 + \sum_i \left[\dfrac{\langle \widehat{H_o^{-1} p_i^2} \rangle}{\langle \widehat{H_o^{-1}} \rangle (p^o_i)^2} - 1\right] P_i^2 = \notag
\\
& = m^2 + P^2 + \sum_i \beta_i P_i^2
\end{align}
where in the second step added and subtracted $P^2$ so as to bring out the Lorentz-invariant part, $m^2 + P^2$, and the correction to it, which is controlled by the three parameters
\begin{align} \label{beta-i}
\beta_i := \dfrac{\langle \widehat{H_o^{-1} p_i^2} \rangle}{\langle \widehat{H_o^{-1}} \rangle (p^o_i)^2} - 1
\end{align}
This expression is not yet satisfactory, because it still involves $p^o_i$ while we would like to see only expectation values of gravitational operators. Since $p^o_i$ is the low-energy limit of the exact solution $p_i(\vec{k})$, the product $p_1^o p_2^o p_3^o$ has to satisfy equation (\ref{god}) in the $k \ll m$ limit, i.e., equation (\ref{class-obs}). This equation is obviously not enough to determine uniquely the unknowns $p_i^o$'s. However, based on reasonable arguments, we can narrow down the range of possible solutions $p_i^o$. The reasoning goes as follows: by definition the $p_i^o$'s do not depend on the $k_i$'s, they can depend only on the expectation values involved in (\ref{god2}). Let us introduce the quantities $\omega_0$ and $\omega_i$
\begin{align}
 \omega_0 := \dfrac{\langle \widehat{H_o^{-1} p_1p_2p_3} \rangle}{\langle \widehat{H_o^{-1}} \rangle}, \qquad \omega_i:= \dfrac{\langle \widehat{H_o^{-1} p_i^2} \rangle}{\langle \widehat{H_o^{-1}} \rangle}
\end{align}
Then we have
\begin{align}
 p_i^o = p_i^o(\omega_0,\omega_1,\omega_2,\omega_3)
\end{align}
Let us now invoke three arguments that impose certain symmetries on the $p_i^o$'s:
\begin{enumerate}
 \item[a)] Since $\omega_0$ is a symmetric quantity with respect to the directions of anisotropy, it is reasonable to assume that the dependence of the three $p_i^o$'s on $\omega_0$ is exactly the same.
 \item[b)] The functions $p_i^o$'s are not expected to differ from each other beyond a simple permutation of the arguments $\omega_j$'s (for $j \neq i$), therefore we can write
 \begin{align}
 p_1^o = F(\omega_0,\omega_1,\omega_2,\omega_3), \ \ \ p_2^o = F(\omega_0,\omega_2,\omega_3,\omega_1), \ \ \ p_3^o = F(\omega_0,\omega_3,\omega_1,\omega_2)
 \end{align}
 where $F$ is so far an arbitrary positive function.
 \item[c)] Each function $p_i^o$ should be symmetric with respect to the remaining $\omega_j, \omega_k$ (for $j \neq i \neq k$).
\end{enumerate}
% These arguments constitute the strongest constraints we can put on the explicit expression of the $p_i^o$'s in terms of the $\omega_j$'s without making ad-hoc choices. 
These conditions do not select a unique form of the $p_i^o$'s; they however imply that, if we assume $p_i^o$ to depend on one of the $\omega_j$'s, then $p_i^o$ must depend on all $\omega_j$'s (otherwise equation (\ref{class-obs}) cannot be satisfied). Therefore we can distinguish two classes of solutions $\{p_i^o\}$: the first class contains solutions that depend only on $\omega_0$, the second class contains solutions that depend on all $\omega_j$'s. It turns out that, by means of condition a), there is only one solution of the first class:
\begin{align}
 p_1^o=p_2^o=p_3^o=\omega_0^\frac{1}{3}
\end{align}
Plugging this into (\ref{beta-i}), we find that the Lorentz-deformation parameters are
\begin{align}
\beta_i = \omega_0^{-\frac{2}{3}} \omega_i - 1 = \dfrac{\langle \widehat{H_o^{-1} p_i^2} \rangle}{\langle \widehat{H_o^{-1} p_1p_2p_3} \rangle^{\frac{2}{3}} \langle \widehat{H_o^{-1}} \rangle^{\frac{1}{3}}} - 1
\end{align}
Interestingly, for this class of dressed metrics the deformation in direction $i$, $\beta_i P_i^2$, depends only on $\omega_0$ and the corresponding $\omega_i$. However, this solution has a major problem: it does not reproduce the Lorentz-invariant dispersion relation in the ``classical gravity'' limit, and hence it is inconsistent with General Relativity. Indeed, if we replace the expectation value of product of operators with the product of expectation values, the parameters $\beta_i$ reduce to
\begin{align}
\beta_i \approx \dfrac{\langle \widehat{H_o^{-1}} \rangle \langle \hat{p}_i \rangle^2}{\langle \widehat{H_o^{-1}} \rangle^{\frac{2}{3}} \langle \hat{p}_1 \rangle^{\frac{2}{3}} \langle \hat{p}_2 \rangle^{\frac{2}{3}} \langle \hat{p}_3 \rangle^{\frac{2}{3}} \langle \widehat{H_o^{-1}} \rangle^{\frac{1}{3}}} - 1 = \dfrac{\langle \hat{p}_i \rangle^2}{\langle \hat{p}_1 \rangle^{\frac{2}{3}} \langle \hat{p}_2 \rangle^{\frac{2}{3}} \langle \hat{p}_3 \rangle^{\frac{2}{3}}} - 1 
\end{align}
which do not vanish in general (unless the semiclassical state of geometry is isotropic, in which case all three expectation values $\langle \hat{p}_i \rangle$ coincide).

We conclude that such solution is not acceptable, and look for solutions in the second class, that is, where each $p^o_i$ depends on all $\omega_j$'s. Making use of the symmetry arguments and an analyticity condition, we can surprisingly derive a unique\footnote{For the details of the conditions and the proof of uniqueness, see appendix \ref{proof2}.}, physically acceptable, and rather simple expression for the $p_i^o$'s:
\begin{align}
p^o_1 = \omega_0^{\frac{1}{3}} \omega_1^\frac{1}{3} (\omega_2 \omega_3)^{-\frac{1}{6}}, \ \ \ \ \ p^o_2 = \omega_0^{\frac{1}{3}} \omega_2^\frac{1}{3} (\omega_3 \omega_1)^{-\frac{1}{6}}, \ \ \ \ \ p^o_3 = \omega_0^{\frac{1}{3}} \omega_3^\frac{1}{3} (\omega_1 \omega_2)^{-\frac{1}{6}}
\end{align}

It is then easy to see that the three parameters $\beta_i$'s are in this case equal: the dispersion relation for particles of wave vector $k_\mu$ propagating on quantum Bianchi I spacetime is therefore given by
\begin{align}\label{Disp1}
E^2 = m^2 + (1 + \beta) P^2
\end{align}
where
\begin{align}
\beta = \dfrac{\langle \widehat{H_o^{-1} p_1^2} \rangle^{\frac{1}{3}} \langle \widehat{H_o^{-1} p_2^2} \rangle^{\frac{1}{3}} \langle \widehat{H_o^{-1} p_3^2} \rangle^{\frac{1}{3}}}{\langle \widehat{H_o^{-1} p_1p_2p_3} \rangle^{\frac{2}{3}} \langle \widehat{H_o^{-1}} \rangle^{\frac{1}{3}}} - 1
\end{align}
While $\beta$ does depend on quantum anisotropies (via the expectation values of $\widehat{H_o^{-1} p_i^2}$, which may be different for different directions $i$'s), it is a unique parameter and hence deforms the dispersion relation in an isotropic fashion. This is rather surprising, considering that our general analysis involved three (in principle different) deformation parameters $\beta_i$.

\begin{figure}
\begin{centering}
\includegraphics[height=1.9in]{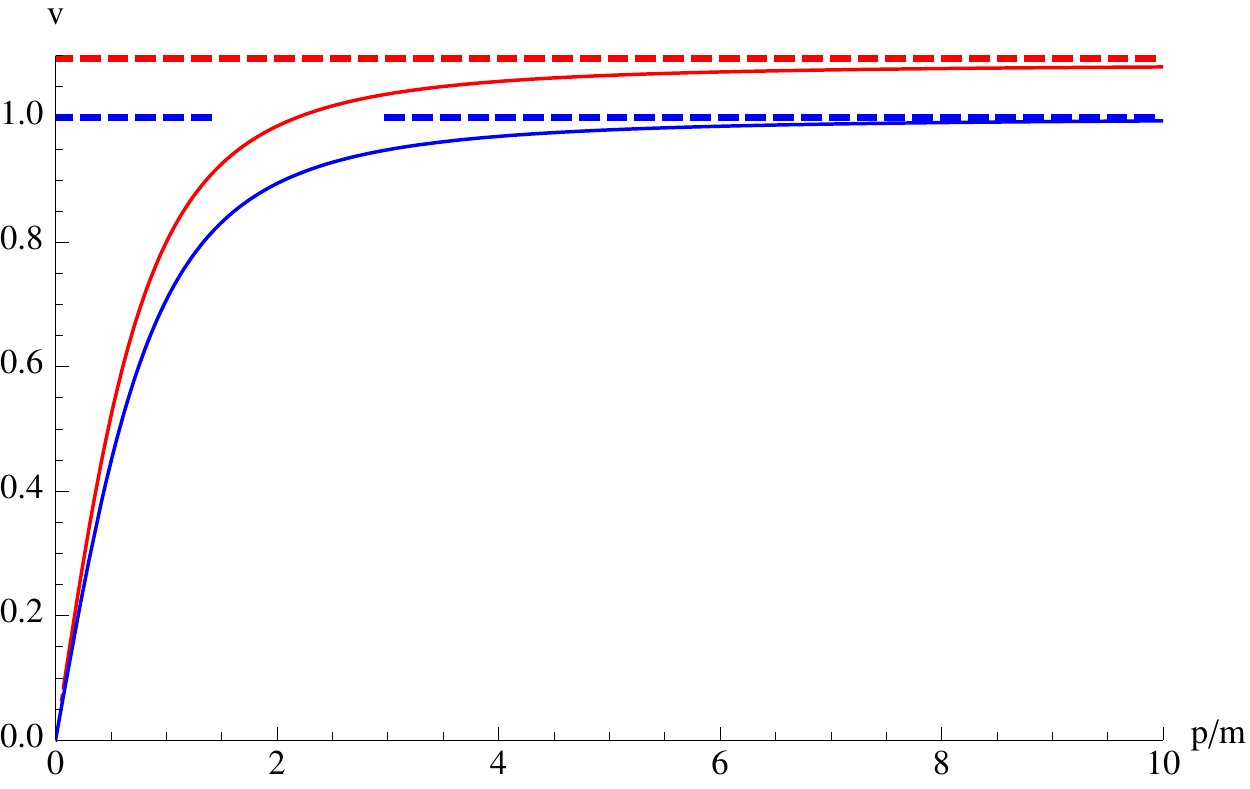}
\caption{{\footnotesize Speed $v = v(P)$ of different modes of the massive field. Red = semiclassical spacetime ($\beta \approx 0$); Blue = quantum spacetime ($\beta \approx 0.2$). The dashed lines represent the speed of light in the two cases.}}\label{fig:vels2}
\par\end{centering}
\end{figure}

It is also easy to check consistency with the isotropic case studied in \cite{rainbow1}. First, assuming that the scale factors operators $\hat a_i$'s commute with each other, the operators $\hat p_i$ are then defined as
\begin{align}
\hat p_1:=\hat a_2 \hat a_3 , \qquad \hat p_2:=\hat a_1 \hat a_3 , \qquad \hat p_3:=\hat a_1 \hat a_2\ .
\end{align}
Following a certain choice of symmetric ordering
\begin{align}
\forall f\in\mathcal{C}_{+}^\infty,\quad &\widehat{H_o^{-1} f(a_i)}:=\sqrt{f(\hat a_i)} \hat H_o^{-1} \sqrt{f(\hat a_i)}\ ,
\end{align}
we evaluate $\beta$ on a state $\Psi_o$ satisfying the condition
\begin{align}
\hat a_1 \Psi_o = \hat a_2 \Psi_o = \hat a_3 \Psi_o\ ,
\end{align}
which is our choice to define an \emph{isotropic state} in the Hilbert space $\mathcal{H}_G$ of Bianchi I geometries\footnote{The consistency between the anisotropic and isotropic cases for $\beta$ can be achieved through a \emph{weaker} definition of quantum isotropy.}. Using the fact that the operators $\hat a_i$ commute with each other, we immediately find that $\beta$ can be written in terms of a single scale factor operator, say $\hat a := \hat a_1$. The explicit form is
\begin{align}
\beta = \dfrac{\langle \widehat{H_o^{-1} a^4} \rangle}{\langle \widehat{H_o^{-1} a^6} \rangle^{\frac{2}{3}} \langle \widehat{H_o^{-1}} \rangle^{\frac{1}{3}}} - 1
\end{align}
This coincides with the deformation parameter computed in \cite{rainbow1} for a state in the Hilbert space of FLRW geometries. Therefore, all results found there pass to the Bianchi I case. In particular, we can use the deformed dispersion relation to compute the speed of propagation of modes of the scalar field, and see how this depends on the (modulus) of momentum (see Figure \ref{fig:vels2}). Notice that, for $\beta \sim 1$ (a highly non-classical situation) the speed of massive particles approaches but never exceeds the (deformed) speed of light (which is greater or less than $1$ depending on the sign of $\beta$). On the other hand, for $\beta \ll 1$ (a semiclassical situation) the plot coincides with the classical one.

		\section{Evaluation of $\beta$ in isotropic LQC}
Considering the importance of the parameter $\beta$ in controlling the deformation of the dispersion relation, we here give an estimation for its value in the context of isotropic loop quantum cosmology (LQC) \cite{LQC1, LQC2}. Let us first recall that, in LQC, the volume operator $\hat{V}$ is well-defined and has continuous positive spectrum, $\mathbb{R}_+$. In light of the evolution dictated by the quantum Hamiltonian $\hat{H}_o$, however, one usually restricts to a superselected sector given by a lattice $L \subset \mathbb{R}_+$. Then, the most general state will be a linear superposition of volume-eigenstates $|v\rangle$, where only values $v \in L$ appear. However, as we are interested in evaluating $\beta$ today, we shall consider a Gaussian state peaked on large volume, $v_o$, with width $s$. Given that the Gaussian-like coefficient $e^{-\ln(v/v_o)^2/4s^2}$ (the gaussianity is in $u := \ln(v)$ rather than $v$ itself, since we recall that $v$ is positive) changes by $e^{-\ln(v/v_o)/2vs^2} - 1 \approx - \ln(v/v_o)/2vs^2$ when we move from the component $v$ to the component $v + 1$ (for large $v$), we can effectively replace the sum over finite steps with an integral over all $\mathbb{R}_+$. Thus, in the end our semiclassical state is of the form
\begin{align} \label{LQC-semiclassical-psi}
| \Psi_{v_o} \rangle = \dfrac{1}{N} \int_0^\infty dv \ e^{-\ln(v/v_o)^2/4s^2} | v \rangle
\end{align}
where $N$ is the normalization constant. It is immediate to check that, for $s$ small enough, this state is indeed peaked on $v_o$ with relative dispersion given by $s$ itself:
\begin{align}
\langle \Psi_{v_o} | \hat{V} | \Psi_{v_o} \rangle =  v_o e^{3 s^2/2} \approx v_o, \ \ \ \ \ \delta V := \sqrt{\dfrac{\langle \Psi_{v_o} | \hat{V}^2 | \Psi_{v_o} \rangle}{\langle \Psi_{v_o} | \hat{V} | \Psi_{v_o} \rangle^2} - 1} = \sqrt{e^{s^2} - 1} \approx s
\end{align}
Now, in the definition of $\beta$ there appear powers of $\hat a \sim \hat V^{1/3}$ and $\hat H_o^{-1}$. While it should technically be possible to compute the action of the latter on the states $|v\rangle$, the result is certainly not analytic, and complicates the evaluation of $\beta$. For the sake of simplicity, we assume that the factors $\hat H_o^{-1}$ can be pulled out of the expectation values, and evaluated independently.\footnote{
Another option is to start from a slightly different system altogether. The presence of $H_o^{-1}$ is due to the fact that the Hamiltonian constraint (\ref{ham-con}) we started with is quadratic in the momentum of our physical clock. If one makes a different choice of time (e.g., irrotational dust \cite{Dust1, Dust2, Dust3}), then such factor would disappear. Alternatively, one may follow \cite{mena}, where the authors implement the quadratic constraint as an operator and then reduce it at the quantum level.
}
Then, one finds
\begin{align}
\beta = \dfrac{\langle \Psi_{v_o} | \hat{V}^{\frac{4}{3}} | \Psi_{v_o} \rangle}{\langle \Psi_{v_o} | \hat{V}^2 | \Psi_{v_o} \rangle^{\frac{2}{3}}} - 1 = e^{-4s^2/9} - 1 \approx - \dfrac{4}{9} s^2
\end{align}
Thus, LQC predicts (approximately) a value $\beta \sim -\delta V^2$ today. The first thing to notice is that this number is extremely small: recall that the meaning of dispersion $\delta V$ is to characterize the semiclassicality of state $\Psi_{v_o}$: the smaller this quantity, the more classical the state is. But today we live in an extremely classical universe, so we should consider the value of $\delta V$ to be very small. Consequently, the value of $\beta$ is \emph{quadratically} small: this is consistent with the experimental observations that Lorentz-symmetry is unbroken, even when observing extremely high-energetic particles (such as GRB). The second thing to notice is the sign of the parameter: it is negative. Based on the discussion of \cite{liberati}, this means that high-energy quanta of the scalar field approach a dressed speed of light which is \emph{lower} than the bare one (at which gravitons supposedly move). Hence -- while this remains an apparent Lorentz-violating effect -- detection is unlikely. In fact, strong experimental bounds exist for particles moving faster than gravitons (based on Cerenkov radiation), but not much can be said for particles moving slower than gravitons. According to \cite{liberati}, the current bounds for this case are $|\beta| \lesssim 10^{-2}$, giving $\delta V \lesssim 10^{-1}$ which fits for a semiclassicality parameter.

Finally, a note about evolution: one might expect that, while today $\beta$ is tiny, in the far past (when the universe was supposedly in a much more ``quantum'' state) such parameter was large. We do not have an explicit computation about the time evolution of $\beta$ (one would have to evaluate it on state $|\Psi_{v_o}(\Phi)\rangle := e^{-i \hat{H}_o \Phi} | \Psi_{v_o} \rangle$, though as already explained this task cannot be performed analytically due to the presence of $\hat H_o$), but several LQC numerical computations confirm that Gaussian-like states of the form considered above remain of this form throughout the evolution. In particular, at primordial times (i.e., close to the big bounce, which in LQC replaces the Big Bang singularity), the state is simply $|\Psi_{v_o}(\Phi)\rangle \approx |\Psi_{v(\Phi)}\rangle$, where $v$ satisfies
\begin{align} \label{friedmann-eff}
\left(\dfrac{\dot{v}}{v}\right)^2 = 3 \kappa \rho \left(1 - \dfrac{\rho}{\rho_{\text{max}}}\right), \ \ \ \ \ \ \ \ \ \ \rho_{\text{max}} \approx 0.41 \ \rho_P\ ,
\end{align}
with the initial condition $v_0:=v(0)$. Since the estimation of $\beta$ in the computation above does not care about the specific value of $v_o$, we conclude that $\beta$ remains (approximately) constant -- and tiny -- throughout the evolution, and in particular it is so in the primordial past. Of course, this result is limited to states of the form (\ref{LQC-semiclassical-psi}), which are in no way unique or preferred within the theory.
		\section{Summary and comments}
In this article we presented a construction of effective cosmological models, valid in the test-field approximation, which describe the propagation of a massive quantum scalar field on a quantum anisotropic cosmological spacetime. The aims were: i) to check the validity and consistency of the approach introduced in \cite{rainbow1} in the anisotropic case, ii) to investigate the modification in the dispersion relation for the scalar field modes in presence of anisotropies and the phenomenology which follows from it, iii) to establish an estimation of the deformation parameter $\beta$ in the case of isotropic cosmology and specifically in the context of loop quantum cosmology.

The construction results in a variety of effective rainbow metrics, in which particles of distinct momenta propagate on slightly different geometries. The dynamics of each scalar field mode on the corresponding effective classical spacetime is equivalent to that of the same mode on the fundamental quantum spacetime. Furthermore, despite the fact that the construction does not lead to a unique effective model, the consistency with the expected semiclassical limit (in addition to imposing certain \emph{partial} analyticity conditions on the parameters determining such anisotropic models) defines a unique deformation parameter $\beta$ in the modified dispersion relation of the modes. As expected, this deformation parameter depends on the quantum anisotropies; however, the deformation it induces is isotropic, which is surprising since our starting point involved three parameters $\beta_i$. Moreover, we approximately estimated the size of this apparent Lorentz-violation for a semiclassical state of geometry defined in LQC, finding it to be completely negligible throughout the entire history of the universe.

Finally, it is not clear whether this deformation would be measurable even in principle, as it simply amounts to a rescaling of the speed of light, and hence -- if $\beta$ is the same for all matter species in the universe -- it would have no physical significance. On the other hand, the situation is different if each matter type had its own parameter $\beta$: for example, in \cite{liberati} it is suggested that gravitons might travel at the bare speed of light, in which case one expects (detectable) Cerenkov effect. As of now we cannot tell for sure, since the only system we explicitly studied is the massive scalar field. Also, while the case considered here is more general than the one studied in \cite{rainbow1}, we are still dealing with finitely many degrees of freedom. It is conceivable that, once the full quantum theory of gravity is taken into account, effective spacetimes would present a richer phenomenology.
\section*{Acknowledgments}
This work has been partially financed by the grants of Polish Narodowe Centrum Nauki nr 2011/02/A/ST2/00300.
\begin{appendices}
		\section{Proof of non-analyticity of the solutions to the system (\ref{fakegod}) at the $0$ point}\label{proof1}
We consider the following system
\begin{align}\label{fakegod1}
\forall i=1,2,3:\ \qquad p_i^2 k_i^2 + \frac{p_1p_2p_3}{3} \ m^2 - \gamma_i(k_i) = 0
\end{align}
with
\begin{align}
\gamma_i(k_i) := \dfrac{\langle \widehat{H_o^{-1} p_i^2} \rangle}{\langle \widehat{H_o^{-1}} \rangle} k_i^2 + \dfrac{\langle \widehat{H_o^{-1} p_1p_2p_3} \rangle}{3\langle \widehat{H_o^{-1}} \rangle} m^2
\end{align}

Let us assume that the functions $p_i(k_1,k_2,k_3)$ are analytic at the point $\{k_i\}=0$. This implies that ${p_i^2 k_i^2}_{|k_{i}=0}=0$. Let us then investigate the solutions of (\ref{fakegod1}) around $0$.

On the one hand, when $k_1=0$, (\ref{fakegod1}) reduces to
\begin{align}
\left\{
\begin{array}{l}
\dfrac{1}{3}p_1p_2p_3 - \dfrac{\langle \widehat{H_o^{-1} p_1p_2p_3} \rangle}{3\langle \widehat{H_o^{-1}} \rangle} = 0 \\
 \\
p_2(0,k_2,k_3)^2 k_2^2 + \frac{1}{3}p_1p_2p_3  \ m^2 - \gamma_2(k_2) = 0 \\
 \\
p_3(0,k_2,k_3)^2 k_3^2 + \frac{1}{3}p_1p_2p_3  \ m^2 - \gamma_3(k_3) = 0
\end{array}
\right.
\end{align}
This system implies that
\begin{align} \label{proof1-1}
p_1(0,k_2,k_3)^2=\dfrac{\langle \widehat{H_o^{-1} p_1p_2p_3} \rangle^2}{\langle \widehat{H_o^{-1} p_2^2} \rangle \langle \widehat{H_o^{-1} p_3^2} \rangle}
\end{align}

One the other hand, when $k_2=0$, (\ref{fakegod1}) reduces to
\begin{align}
\left\{
\begin{array}{l}
\dfrac{1}{3}p_1p_2p_3 - \dfrac{\langle \widehat{H_o^{-1} p_1p_2p_3} \rangle}{3\langle \widehat{H_o^{-1}} \rangle} = 0 \\
 \\
p_1(k_1,0,k_3)^2 k_1^2 + \frac{1}{3}p_1p_2p_3  \ m^2 - \gamma_1(k_1) = 0 \\
 \\
p_3(k_1,0,k_3)^2 k_3^2 + \frac{1}{3}p_1p_2p_3  \ m^2 - \gamma_3(k_3) = 0
\end{array}
\right.
\end{align}
which implies that 
\begin{align} \label{proof1-2}
p_1(k_1,0,k_3)^2=\dfrac{\langle \widehat{H_o^{-1} p_1^2} \rangle}{\langle \widehat{H_o^{-1}} \rangle}
\end{align}

Therefore, thanks to the analyticity of $p_1$ at $\{k_i\}=0$, from (\ref{proof1-1}) and (\ref{proof1-2}) we obtain
\begin{align}
\dfrac{\langle \widehat{H_o^{-1} p_1^2} \rangle}{\langle \widehat{H_o^{-1}} \rangle}=\lim_{k_1\rightarrow 0} p_1(k_1,0,k_3)^2=\lim_{k_2\rightarrow 0} p_1(0,k_2,k_3)^2=\dfrac{\langle \widehat{H_o^{-1} p_1p_2p_3} \rangle^2}{\langle \widehat{H_o^{-1} p_2^2} \rangle \langle \widehat{H_o^{-1} p_3^2} \rangle}
\end{align}
which implies that
\begin{align}
\dfrac{\langle \widehat{H_o^{-1} p_1^2} \rangle}{\langle \widehat{H_o^{-1}} \rangle}\dfrac{\langle \widehat{H_o^{-1} p_2^2} \rangle}{\langle \widehat{H_o^{-1}} \rangle}\dfrac{\langle \widehat{H_o^{-1} p_3^2} \rangle}{\langle \widehat{H_o^{-1}} \rangle}=\dfrac{\langle \widehat{H_o^{-1} p_1p_2p_3} \rangle^2}{ \langle \widehat{H_o^{-1}} \rangle^2}
\end{align}
This equation is not verified for a generic state of geometry as we are considering here, thus we showed that the analyticity of the solutions to the system (\ref{fakegod1}) at the $0$ point imposes a non trivial and generally invalid condition on the state of geometry. Requiring analytic $p_i$'s as functions of the modes $k_j$'s at point $0$ (and hence analytic effective scale factors) forces us to discard the system (\ref{fakegod1}) as a valid approach to construct effective rainbow metrics.
		\section{Proof of uniqueness of the physically admissible solution to (\ref{god})}\label{proof2}
We want to determine the expression of $p_i^o$'s in terms of the $\omega_i$'s such that
\begin{align} \label{class-obs1}
p^o_1p^o_2p^o_3 = \omega_0\ .
\end{align}
We can write
\begin{align}
 p_i^o= p_i^o(\omega_0,\omega_1,\omega_2,\omega_3)= \omega_0^\frac{1}{3} \frac{p_i^o(\omega_0,\omega_1,\omega_2,\omega_3)}{\omega_0^\frac{1}{3}}=:\omega_0^\frac{1}{3} \bar p_i^o(\omega_0,\omega_1,\omega_2,\omega_3)\ ,
\end{align}
where $\bar p_i^o$ is now a dimensionless quantity. We can go further and write
\begin{align}
\bar p_i^o(\omega_0,\omega_1,\omega_2,\omega_3)=:\tilde p_i^o(x_0,x_1,\dots)\ ,
\end{align}
where the variables $x_\alpha$, with $\alpha$ an integer label, are dimensionless and have the form
\begin{align}
x_\alpha=\omega_0^{r_\alpha}\omega_1^{s_\alpha}\omega_2^{p_\alpha}\omega_3^{q_\alpha}\ l_p^{2(-3 r_\alpha-2 s_\alpha-2 p_\alpha-2 q_\alpha)},
\end{align}
with $r_\alpha$, $s_\alpha$, $p_\alpha$ and $q_\alpha$ being arbitrary real numbers and $l_p$ Planck length. Notice that we have the following property
\begin{align}\label{xprop}
\forall\ n, m \in \mathbbm{N},\ \forall\ \{i_\alpha\} \in \mathbbm{N}^n ,\ \exists\ x_\gamma\ :\ \prod_{\alpha=m}^{n+m-1} x_\alpha^{i_\alpha}= x_\gamma\ .
\end{align}

{\it Assuming that $\tilde p_i^o$ is analytic in the $x_\alpha$'s}, then using the symmetry arguments mentioned above and the property (\ref{xprop}), the function $(\tilde p_1^o)^2$ at every point can take the form
\begin{align}
 (\tilde p_1^o)^2=\sum_\alpha c_\alpha\ l_p^{2(-3 r_\alpha-2 s_\alpha-2 p_\alpha-2 q_\alpha)} \left[ \omega_0^{r_\alpha}\omega_1^{s_\alpha}\omega_2^{p_\alpha}\omega_3^{q_\alpha} +\omega_0^{r_\alpha}\omega_1^{s_\alpha}\omega_2^{q_\alpha}\omega_3^{p_\alpha}\right]\ ,
\end{align}
where $c_\alpha$'s are arbitrary real numbers.

Looking at $\beta_1$, we obtain that in the semiclassical limit $(\tilde p_1^o)^2$ must verify
\begin{align}
\omega_0^\frac{2}{3} (\tilde p_1^o)^2=\omega_1\ ,
\end{align}
which implies that
\begin{align}\label{m1}
\sum_\alpha c_\alpha\ l_p^{2(-3 r_\alpha-2 s_\alpha-2 p_\alpha-2 q_\alpha)} \left[ \omega_0^{r_\alpha+\frac{2}{3}}\omega_1^{s_\alpha-1}\omega_2^{p_\alpha}\omega_3^{q_\alpha} +\omega_0^{r_\alpha+\frac{2}{3}}\omega_1^{-p_\alpha-q_\alpha-\frac{3}{2}r_\alpha-1}\omega_2^{q_\alpha}\omega_3^{p_\alpha}\right]=1\ ,
\end{align}
and hence in each term of the sum with $c_\alpha\neq 0$, the total power (roughly) of each $\hat p_i$ obtained from all the $\omega_i$'s must vanish. This means that for each $\alpha$ we have
\begin{align}
 &\nonumber r_\alpha+\frac{2}{3}+2 s_\alpha-2=0\\
 & r_\alpha+\frac{2}{3}+2 p_\alpha=0\\
 &\nonumber r_\alpha+\frac{2}{3}+2 q_\alpha=0\ ,
\end{align}
and 
\begin{align}\label{ca}
2\sum_\alpha c_\alpha=1\ .
\end{align}
leading to
\begin{align}\label{dimsol}
s_\alpha-1= p_\alpha=q_\alpha=-\frac{r_\alpha}{2}-\frac{1}{3}\qquad \text{and}\qquad 2\sum_\alpha c_\alpha=1\ .
\end{align}

Consequently we obtain
\begin{align}
 (\tilde p_1^o)^2&=2\sum_\alpha c_\alpha\ \omega_0^{-\frac{2}{3}-2q_\alpha}\omega_1^{q_\alpha+1}\omega_2^{q_\alpha}\omega_3^{q_\alpha}\\
 &=:\omega_1 G(\omega_0,\omega_1,\omega_2,\omega_3),
\end{align}
where the Planck length factors disappeared from the expression of $p_1$ as the power of the Planck length in each of the factors vanishes thanks to the first equation in \ref{dimsol}.\\

Similarly, we get
\begin{align}
(\tilde p_2^o)^2&=2\sum_\alpha c_\alpha\ \omega_0^{-\frac{2}{3}-2q_\alpha}\omega_2^{q_\alpha+1}\omega_1^{q_\alpha}\omega_3^{q_\alpha}=\omega_2 G(\omega_0,\omega_1,\omega_2,\omega_3)\\
 \nonumber  (\tilde p_3^o)^2&=2\sum_\alpha c_\alpha\ \omega_0^{-\frac{2}{3}-2q_\alpha}\omega_3^{q_\alpha+1}\omega_1^{q_\alpha}\omega_2^{q_\alpha}=\omega_3 G(\omega_0,\omega_1,\omega_2,\omega_3)\ .
\end{align}

Now, from (\ref{class-obs1}) we have that the $\tilde p_i^o$'s must verify
\begin{align}
(\tilde p_1^o)^2(\tilde p_2^o)^2(\tilde p_3^o)^2 = 1\ ,
\end{align}
meaning that
\begin{align}
\omega_1 \omega_2 \omega_3\ G(\omega_0,\omega_1,\omega_2,\omega_3)^3=1\ .
\end{align}
Hence
\begin{align}
G(\omega_0,\omega_1,\omega_2,\omega_3)=(\omega_1 \omega_2 \omega_3)^{-\frac{1}{3}}\ .
\end{align}
It then follows that
\begin{align}
 \nonumber (\tilde p_1^o)^2&=\omega_1^{\frac{2}{3}}\omega_2^{-\frac{1}{3}}\omega_3^{-\frac{1}{3}}\\
 (\tilde p_2^o)^2&=\omega_2^{\frac{2}{3}}\omega_1^{-\frac{1}{3}}\omega_3^{-\frac{1}{3}}\\
 \nonumber (\tilde p_3^o)^2&=\omega_3^{\frac{2}{3}}\omega_1^{-\frac{1}{3}}\omega_2^{-\frac{1}{3}}\ ,
\end{align}
and therefore
\begin{align}
 \nonumber p_1^o&=\omega_0^{\frac{1}{3}}\omega_1^{\frac{1}{3}}\omega_2^{-\frac{1}{6}}\omega_3^{-\frac{1}{6}}\\
 p_2^o&=\omega_0^{\frac{1}{3}}\omega_2^{\frac{1}{3}}\omega_1^{-\frac{1}{6}}\omega_3^{-\frac{1}{6}}\\
 \nonumber p_3^o&=\omega_0^{\frac{1}{3}}\omega_3^{\frac{1}{3}}\omega_1^{-\frac{1}{6}}\omega_2^{-\frac{1}{6}},
\end{align}
which is the final solution analyzed in section \ref{section4}.

\end{appendices}

%%%%%%%%%%%%%%%%%%%%%%%%%%%%%%%%%%%%%%%%%%%%%%%%%%%%%%%%%%%%%%%%%%%%%%%%%%%%%%%%%%%%%


\begin{thebibliography}{99}
%%%%%%%%%%%%%%%%%%%%%%%%%%%%%%%%%%%%%%%%%%%%%%%%%%%%%%%%%%%%%%%%%%%%%%%%%%%%%%%%%%%%%

\bibitem{rainbow1} M. Assanioussi, A. Dapor, J. Lewandowski - \emph{Rainbow metric from quantum gravity}, Phys. Lett. B \textbf{751}, Pages 302-305 (2015).

\bibitem{emerg1} C. Barcelo, M. Visser, S. Liberati - \emph{Einstein gravity as an emergent phenomenon?}, Int. J. Mod. Phys. D \textbf{10}, 799-806 (2001).

\bibitem{emerg2} D. Oriti - \emph{Group field theory as the microscopic description of the quantum spacetime fluid: A New perspective on the continuum in quantum gravity}, PoS QG-PH 030 (2007).

\bibitem{QT-FRW} A. Ashtekar, W. Kaminski, J. Lewandowski - \emph{Quantum field theory on a cosmological, quantum spacetime}, Phys. Rev. D \textbf{79}, 064030 (2009).

\bibitem{QT-pert} I. Agullo, A. Ashtekar, W. Nelson - \emph{A Quantum Gravity Extension of the Inflationary Scenario}, Phys. Rev. Lett. \textbf{109}, 251301 (2012).

\bibitem{QT-particles} Y. Tavakoli, J. C. Fabris - \emph{Quantum effects of massive modes in a cosmological quantum space-time}, [arXiv:1511.08823].

\bibitem{QT-BH} R. Gambini, J. Pullin - \emph{Hawking radiation from a spherical loop quantum gravity black hole}, Class. Quant. Grav. \textbf{31}, 115003 (2014). 

% \bibitem{rainbow1} M. Assanioussi, A. Dapor, J. Lewandowski - \emph{Rainbow metric from quantum gravity}, Phys. Lett. B \textbf{751}, Pages 302-305 (2015).

\bibitem{alex} A. Stottmeister, T. Thiemann - \emph{Coherent states, quantum gravity and the Born-Oppenheimer approximation, I: General considerations}, [arXiv:1504.02169].

\bibitem{liberati} R. G. Torrom\'e, M. Letizia, S. Liberati - \emph{Phenomenology of effective geometries from quantum gravity}, Phys. Rev. D \textbf{92}, 124021 (2015).

\bibitem{RAIN-intro1} R. Lafrance, R. C. Myers - \emph{Gravity’s rainbow: Limits for the applicability of the equivalence principle}, Phys. Rev. D \textbf{51}, 2584 (1995).

\bibitem{RAIN-intro2} J. Magueijo, L. Smolin - \emph{Gravity's rainbow}, Class. Quant. Grav. \textbf{21}, 1725-1736 (2004).

\bibitem{Dust1} V. Husain, T. Pawlowski - \emph{Time and a physical Hamiltonian for quantum gravity}, Phys. Rev. Lett. \textbf{108}, 141301 (2012).

\bibitem{Dust2} K.~Giesel and T.~Thiemann - \emph{Scalar Material Reference Systems and Loop Quantum Gravity}, Class. Quant. Grav. {\bf 32}, 135015 (2015).

\bibitem{Dust3} J\c{e}drzej \'Swie\.zewski - \emph{On the properties of the irrotational dust model}, Class. Quant. Grav. \textbf{30}, 237001 (2013).

\bibitem{mena} L. Castell\'o Gomar, M. Mart\'in-Benito, G. A. Mena Marug\'an - \emph{Gauge-Invariant Perturbations in Hybrid Quantum Cosmology}, JCAP \textbf{06}, 045 (2015).

\bibitem{LQC1} A. Ashtekar - \emph{An Introduction to Loop Quantum Gravity through Cosmology}, Nuovo Cimento 112B, 1-20 (2007).

\bibitem{LQC2} A. Ashtekar, A. Corichi, P. Singh - \emph{Robustness of key features of loop quantum cosmology}, Phys. Rev. D \textbf{77}, 024046 (2008).

\end{thebibliography}
\end{document}